\documentstyle[12pt]{article}

\newcommand{\beq}[1]{\begin{equation} \label{#1} }
\newcommand{\eeq}    {\end{equation}}
\newcommand{\nuc}[2]{\mbox{$^{#1}{\rm #2}$}}

\begin{document}
\baselineskip=0.8cm
\pagenumbering{roman}

\mbox{}

\noindent{\large Russian Research Center "Kurchatov
Institute"}

\noindent{\large Preprint IAE-5670/2}

\vspace{1cm}

\noindent{\large A.L.Barabanov}
\vspace{1cm}

\noindent{\large FISSION FRAGMENT ORIENTATION AND}

\noindent{\large $\gamma$ RAY EMISSION ANISOTROPY}
\vspace{1cm}

\noindent\mbox{Report at the 12-th Meeting}

\noindent\mbox{on Physics of Nuclear Fission,}

\noindent\mbox{September 27-30, 1993, Obninsk, Russia}
\vspace{3cm}

\begin{center}
{\large Moscow -- 1993}
\end{center}

\pagebreak

\baselineskip=0.5cm

\mbox{}

\noindent Key words: ternary fission, $\gamma$ ray emission,
angu\-lar dis\-tri\-bu\-tion,
spin-tensor of orientation

\mbox{}

Experimental data on  angular  distributions  of  $\gamma$
rays  emit\-ted  from  binary  and  ternary  spontaneous
fission  of \nuc{252}{Cf}
 are  ana\-ly\-zed.  Their difference
indicates that the alignment of fragments is higher in
ternary fission than in binary one.  The  conse\-quen\-ces
of possible relation between the mechanism of  ter\-na\-ry
fission and the excitation of collective modes  during
the saddle - to - scission stage are discussed.

\pagebreak

\section{Introduction}
\pagenumbering{arabic}

   The descent of the fissioning nucleus  from  saddle
to scission point is of interest as the fragment mass,
charge, excitation energy and spin distributions   are
formed at this  stage.  It  is  well  known  that  the
fragment spins are relatively high (\mbox{$\langle J \rangle
\sim  7-8 $})  even
for the spontaneous fission of zero spin nucleus  like
\nuc{252}{Cf}. Besides the fragment spins are aligned  in  the
plane per\-pen\-di\-cu\-lar to the  fission  axis  causing  an
anisotropy of $\gamma$ rays emitted from fragments. The  most
probable reason for fragment spins and their alignment
is the excitation of collective vibrational modes like
bending or wriggling at the saddle - to - scission  stage.
Although such modes are being  discussed  for  a  long
time \cite{1,2} their possible influence on  the  formation
of mass and energy distributions of fragments  is  not
really  taken  into  account  (see,  e.g.,  Refs. \cite{3,4}).
Similarly, vibrational modes are ignored in the A.Bohr
model \cite{5} for fission  fragment  angular  distribution
since they violate the axial  symmetry  of  fissioning
nucleus at the descent stage and thus can perturb  the
dis\-tri\-bu\-tion of $K$ quantum number formed at the  saddle
point.

   Recently  the  new  experimental  data  on  angular
dis\-tri\-bu\-tions  of   prompt   $\gamma$   rays   emitted   from
spontaneous fission of    \nuc{252}{Cf} have  been  published  by
Pilz and Neubert \cite{6}. For the first time  the  results
were obtained  for  ternary  fission  of \nuc{252}{Cf}  which
probability is $\sim 1/300$ of binary fission\cite{7}.  Earlier
only  the   an\-gu\-lar   ani\-so\-tro\-py
\mbox{$w^{(t)}(0^{0})/w^{(t)}(90^{0})=1.015 \pm 0.022$}
for ter\-na\-ry fis\-sion  of \nuc{252}{Cf} had  been
mea\-sured \cite{8};  this  re\-sult  was  in\-ter\-pre\-ted
as  the
evidence for the  destruction  of  fragment  alignment
owing to the $\alpha$ particle emission. Although  the  value
of angular anisotropy in ternary fission was confirmed
in Ref. \cite{6} the measured angular distributions were found
to be anisotropic! Whereas the  angular  distributions
\mbox{$w^{(b)}(\theta)$} in binary fission have the maximums at
$0^{0}$ and  $180^{0}$  with respect to the light fragment
momentum the
distributions  \mbox{$w^{(t)}(\theta)$}  in  ternary   fission
 reveal
maximums at $20-40^{0}$   and  $140-160^{0}$ .  The  distributions
\mbox{$w^{(b)}(\theta)$} and \mbox{$w^{(t)}(\theta)$}
 are  presented  in  Ref. \cite{6}  for  8
groups of $\gamma$ rays with energies  from  151-242  keV  to
978-1208 keV, but the  energy  dependence  of  angular
distributions is quite week.

   Pilz and Neubert interpreted their results  as  the
evidence for the tilting of the fragment alignment  by
the recoiling $\alpha$ particle  causing  the  shift  of  the
maximums of $\gamma$ ray angular distributions  from  $0^{0}$   and
$180^{0}$  to $20-40^{0}$  and $140-160^{0}$. It seems,  however,
 that
the accidental tiltings would widen the  maximums  but
not shift  them.  The  purpose  of  this  work  is  to
demonstrate  the  possibility   of   more   consistent
explanation of the results obtained in Ref. \cite{6} based  on
the standard description of the  angular  distribution
of $\gamma$ rays emitted from aligned nuclei \cite{9}.

\section{Helicity representation in fission}

   The reason for angular anisotropy of prompt $\gamma$  rays
is the alignment of fragment spins. Choosing the light
fragment mo\-men\-tum for the direction of the $z$  axis  we
can write in  the  center  of  mass  system  the  wave
function of two separated light  and  heavy  fragments
with spins $J_{1}$  and $J_{2}$  in the form of the superposition
\beq{1}
\psi_{J_{1}J_{2}}=\sum_{K_{1}K_{2}}g(K_{1},K_{2})
\psi_{J_{1}K_{1}}\psi_{J_{2}K_{2}},\;\;
\sum_{K_{1}K_{2}}\mid g(K_{1},K_{2})\mid^{2}=1,
\eeq
where $K_{1}$ is the projection of spin $J_{1}$ to the $z$ axis or
to the momentum $\vec{p}_{1}$  of the light fragment, that is  the
helicity of the light fragment; $K_{2}$  is  the  projection
of spin $J_{2}$  to the $z$ axis or to  the  momentum
$\vec{p}_{1}=-\vec{p}_{2}$,
that is the helicity of the heavy  fragment  with  the
opposite sign (for simplicity we  shall  name  $K_{2}$   the
helicity of the heavy fragment); and $g(K_{1},K_{2})$ are  the
amplitudes  of  fission  in  helicity   representation
introduced in Ref.\cite{10}. After  the  averaging  over  the
ensemble  of  fissioning  nuclei  and  summation  over
helicity of additional  fragment  the  spin  state  of
light fragment is described by the density matrix
\beq{2}
\rho_{K_{1}K'_{1}}=\sum_{K_{2}}\overline{g(K_{1},K_{2})
g(K'_{1},K_{2})^{\ast}},\;\;\;\;
\sum_{K_{1}} \rho_{K_{1}K_{1}}=1,
\eeq
or by the set of spin-tensors of orientation
\beq{3}
\tau_{Qq}(J_{1})=\sum_{K_{1}K'_{1}}C_{J_{1}K_{1}Qq}^{J_{1}K'_{1}}
\rho_{K_{1}K'_{1}},\;\;\;\;
\tau_{00}(J_{1})=1.
\eeq
Due to the axial symmetry with respect to the  fission
direction for spontaneous fission of    \nuc{252}{Cf}  with  zero
spin we obtain for spin-tensors
\beq{4}
\tau_{Qq}(J_{1})=\tau_{Qq}(J_{1})\delta_{q0}.
\eeq
The  spin-tensors  of  fragment  orientation  for  the
fission of pre\-li\-mi\-na\-ry oriented nuclei were calculated
in Ref.\cite{10}. As the consequence of parity conservation
\beq{5}
\mid g(K_{1},K_{2})\mid^{2}=\mid g(-K_{1},-K_{2})\mid^{2},
\eeq
thus $\tau_{Q0}(J_{1}) \not= 0$ only for $Q=0,2,4 \ldots$
$(Q < 2J_{1})$.  By  the
same  way  the  set  of  spin-tensors  $\tau_{Q0}(J_{2}) \not=0$
for $Q=0,2\ldots 2J_{2}$  describes the  spin  state  of  the
 heavy fragment.

   The total helicity $K_{1}+K_{2}$ is nothing  else  but  the
projection $K$ of the spin $J$ of  fissioning  nucleus  to
the fission axis. Indeed, the total angular momentum $\vec{J}$
of the fissioning system remains  unvarying,  thus  we
have after the scission
\beq{6}
\vec{J}=\vec{J}_{1}+\vec{J}_{2}+\vec{L},
\eeq
where $\vec{L}$ is  the  fragment  orbital  angular  momentum.
Projecting this  equation  to  the  fission  axis  and
taking into account that the  angular  momentum  $\vec{L}$  is
perpendicular to the mo\-men\-ta \mbox{$\vec{p}_{1}=-\vec{p}_{2}$},
we  get  $K=K_{1}+K_{2}$.
The dependence of fission  probability  on  the  total
helicity
\beq{7}
\beta_{K}=\sum_{K_{1}}\mid g(K_{1},K-K_{1}\mid^{2},\;\;\;\;
\sum_{K}\beta_{K}=1
\eeq
is the significant characteristic of the process since
its shape determines the form of the fission  fragment
angular distribution \cite{11} (see also Refs.\cite{12,10})
\begin{eqnarray}
&&w(\vec{n}_{f})=\sum_{Qq}
\left(\frac{2Q+1}{4\pi}\right)^\frac{1}{2}
\tau_{Qq}(J) \, \alpha_{Q}(J) \,
Y_{Qq}^{\ast}(\vec{n}_{f}), \nonumber \\
&&\oint d\Omega_{f} \, w(\vec{n}_{f})=1, \label{8}
\end{eqnarray}
\beq{9}
\alpha_{Q}(J)=\sum_{K} C_{JKQ0}^{JK} \, \beta_{K},
\eeq
here $\vec{n}_{f}=\vec{p}_{1}/p_{1}$ is the unit  vector  along
 the fission
axis. For the  first  time  the  same  expression  for
angular distribution $w(\vec{n}_{f})$ had been obtained by A.Bohr
\cite{5} on the  assumption  that  it  coincides  with  the
distribution of  orientation  of  nuclear  de\-for\-ma\-tion
axis at the saddle point or on the assumption that the
dis\-tri\-bu\-tion $\beta_{K}$  of total helicity forms at
the  saddle
point and remains unvarying during the descent to  the
scission point. In reality as it was noted  above  the
dependence $\beta'_{K'}$ of  the  fission  probability  on  the
projection $K'$ of the spin $J$ to the deformation axis at
the saddle point may be distorted at the descent stage
owing to the nonaxiality of  the  exciting  collective
modes, therefore
\beq{10}
\beta_{K}=\beta'_{K}+\sum_{K'} f_{KK'}\beta'_{K'},
\eeq
where $f_{KK'}$ are the distortion factors.

   In the spontaneous fission of \nuc{252}{Cf} with  spin $J=0$
the total he\-li\-ci\-ty $K$ is equal to zero thus there is no
problem of conservation  of  distribution  $\beta_{K}$ at the
descent stage. The he\-li\-ci\-ty distributions of the light
and heavy fragments
\beq{11}
\gamma_{K_{1}}=\mid g(K_{1},-K_{1})\mid^{2},\;\;
\gamma_{K_{2}}=\mid g(-K_{2},K_{2})\mid^{2},\;\;
\sum_{K_{j}} \gamma_{K_{j}}=1
\eeq
coincide with each other (although the spins $J_{1}$ and $J_{2}$
may differ by the value of orbital angular momentum $L$)
and are determined completely by the manner of nuclear
motion at the saddle-to-scission stage. We  have  for
spin-tensors of second and fourth ranks  the  explicit
expressions
\beq{12}
\tau_{20}(J)\!=\!\left(\!\frac{J(J\!\!+\!\!1)}
{(2J\!\!-\!\!1)(2J\!\!+\!\!3)}\!\right)
^\frac{1}{2}
\left(\!3\frac{\langle K^{2} \rangle}{J(J\!\!+\!\!1)}\!-\!1\!\right)\!,
\eeq
\begin{eqnarray}
&&\tau_{40}(J)\!=\!
\left(\!\frac{J^{3}(J\!\!+\!\!1)^3}
{(2J\!\!-\!\!3)(2J\!\!-\!\!2)(2J\!\!-\!\!1)
(2J\!\!+\!\!3)(2J\!\!+\!\!4)(2J\!\!+\!\!5)}\!\right)
^\frac{1}{2}\!\!\!\cdot \nonumber \\
&&\cdot \left(\!\!35\frac{\langle K^{4} \rangle}
{J^{2}(\!J\!\!+\!\!1\!)^{2}}
\!-\!30\frac{\langle K^{2} \rangle}{J(\!J\!\!+\!\!1\!)}\!
\left(\!\!1\!\!-\!\!\frac{5}{6J(\!J\!\!+\!\!1\!)}\!\!\right)\!+\!
3\!\left(\!\!1\!\!-\!\!\frac{2}{J(\!J\!\!+\!\!1\!)}\!\!
\right)\!\!\right)\!, \label{13}
\end{eqnarray}
where
\beq{14}
\langle K^{n} \rangle=\sum_{K} K^{n}\gamma_{K}.
\eeq
The bending mode leads obviously to the  distributions
$\gamma_{K_{j}}$ grou\-ped around low values of $K_{j}$ (the
spins $J_{j}$ are
aligned at the  plane  perpendicular  to  the  fission
axis), therefore
\beq{15}
\langle K_{j}^{2} \rangle < \frac{J_{j}(J_{j}+1)}{3},\;\;\;\;
\tau_{20}(J_{j}) < 0,\;\;\;\;j=1,2\;,
\eeq
but the twisting mode  would  populate  the  substates
with high helicities (the spins $J_{j}$  are oriented  along
the fission axis), \linebreak therefore
\beq{16}
\langle K_{j}^{2} \rangle > \frac{J_{j}(J_{j}+1)}{3},\;\;\;\;
\tau_{20}(J_{j}) > 0,\;\;\;\;j=1,2\;.
\eeq
The magnitude of spin-tensor $\tau_{40}(J_{j})$ of fourth rank is
de\-ter\-mi\-ned by the more subtle characteristics  of  the
distribution $\gamma_{K_{j}}$.

\section{$\gamma$ ray emission from aligned nuclei}

   According to the standard formalism \cite{9} the angular
dis\-tri\-bu\-tion of $\gamma$ rays of multipolarity $L$
emitted from
the aligned nucleus with spin $J_{i}$  in its transition  to
the state with spin $J_{f}$  is of the form
\begin{eqnarray}
&&w(\vec{n}_{\gamma})=\frac{1}{4\pi}\sum_{Q=0,2\ldots}
(2Q+1)\,C_{L1Q0}^{L1}\,U(J_{f}LJ_{i}Q,J_{i}L)\cdot \nonumber \\
&&\;\;\;\;\;\;\;\;\;\;\;\;\;\;\;\;\;\;\;\;\;\;\;\;\;\;\;\;\;\;
\cdot\,\tau_{Q0}(J_{i})\,P_{Q}(\cos \theta_{\gamma}), \nonumber \\
&&\oint d\Omega_{\gamma}\,w(\vec{n}_{\gamma})=1, \label{17}
\end{eqnarray}
where $\theta_{\gamma}$ is the angle between the unit
vector $\vec{n}_{\gamma}$ along
the $\gamma$ ray momentum and the axis of nuclear  alignment,
$U(abcd,ef)$ is  the  normalized  Racah  function  \cite{13},
$P_{Q}(\cos \theta_{\gamma})$ are Legendre polynomials
\begin{eqnarray}
&&P_{2}(\cos \theta)=\frac{1}{2}(3\cos^{2} \theta-1), \nonumber \\
&&P_{4}(\cos \theta)=\frac{1}{8}(35\cos^{4} \theta
-30\cos^{2} \theta +3), \ldots \label{18}
\end{eqnarray}
The excited fission fragments emit mainly  $\gamma$  rays  of
mul\-ti\-po\-la\-ri\-ties  $e1$,  $m1$  and  $e2$,
so  $L < 2$   and,
therefore, the  angular  distribution  $w(\vec{n}_{\gamma})$
consists
only of the terms corresponding to $Q=0,2$ and $4$.

   The mean number of $\gamma$ rays  emitted  in  spontaneous
fission of \nuc{252}{Cf} is $\langle N_{\gamma} \rangle =9.35$
\cite{14}.  The  deexcitation
process of fission fragment after neutron emission  is
going via two stages (see Fig.\ref{Fig.1}). At first the nucleus
\begin{figure}
\vspace{13cm}
\caption{The  scheme  of  deexcitation   of   even-even
deformed fragment.
\label{Fig.1} }
\end{figure}
emits statistical $\gamma$ rays of $e1$ or  $m1$  multipolarities
and falls in one of the  yrast-line  states,  then  it
descents to the ground state in series of  transitions
between the yrast-line states. In each transition
$J_{i} \rightarrow J_{f}$ the spin-tensors of nuclear
orientation decrease
\beq{19}
\tau_{Q0}(J_{f})=U(LJ_{f}J_{i}Q,J_{i}J_{f})\tau_{Q0}(J_{i}).
\eeq
The factor of decrease, for  example,  for  quadrupole
transition $J_{i} \rightarrow J_{f}=J_{i}\!-\!2$ is
\begin{eqnarray}
&&U(2\,J\!\!-\!\!2\,J\,Q,J\,J\!\!-\!\!2)\!=\!
\frac{1}{4J(J\!\!-\!\!1)(2J\!\!-\!\!1)}\!
\left(\!\frac{A(J,Q)B(J,Q)}
{(2J\!\!-\!\!3)(2J\!\!+\!\!1)}\!\right)^\frac{1}{2}\!,
\nonumber \\
&&A(J,Q)\!=\!(2J\!+\!Q\!+\!1)(2J\!+\!Q)(2J\!+\!Q\!-\!1)
(2J\!+\!Q\!-\!2), \nonumber \\
&&B(J,Q)\!=\!(2J\!-\!Q)(2J\!-\!Q\!-\!1)(2J\!-\!Q\!-\!2)
(2J\!-\!Q\!-\!3). \label{20}
\end{eqnarray}

   It is easy to show that the angular  anisotropy  is
caused  mainly  by  the   stretched   $e2$ transitions
(experimental evidences for this fact were obtained in
Ref.\cite{15}). The Racah functions entering in Eq.\ref{17}
 for $L=2$ and $J_{f}=J_{i}-2$ are positive and are
of the form ($Q=2,4$)
\beq{21}
U(J_{f}2J_{i}2,J_{i}2)=\left(\frac{2(J_{i}+1)(2J_{i}+3)}
{7J_{i}(2J_{i}-1)}\right)^\frac{1}{2},
\eeq
\beq{22}
U(J_{f}2J_{i}4,J_{i}2)=\frac{1}{6}\left(
\frac{2(J_{i}+1)(J_{i}+2)(2J_{i}+3)(2J_{i}+5)}
{7J_{i}(J_{i}-1)(2J_{i}-1)(2J_{i}-3)}\right)^\frac{1}{2}.
\eeq
At the same time the statistical dipole transitions go
to the states with spins $J_{f}=J_{i},J_{i}\pm1$, for
which we have ($Q=2$)
\begin{eqnarray}
U(J_{f}1J_{i}2,J_{i}1)&\!=\!&
\frac{1}{\left(10J_{i}(J_{i}+1)(2J_{i}+3)(2J_{i}-1)\right)
^\frac{1}{2}} \cdot \nonumber \\
\mbox{} &\! \cdot \!&
\left\{ \begin{array}{ll}
        J_{i}(2J_{i}-1),&\mbox{if $J_{f}=J_{i}+1$}\,; \\
        (3-4J_{i}(J_{i}+1)),&\mbox{if $J_{f}=J_{i}$}\,; \\
        (J_{i}+1)(2J_{i}+3),&\mbox{if $J_{f}=J_{i}-1$}\,;
       \end{array} \right.
\label{23}
\end{eqnarray}
so the angular anisotropy in the transition
$J_{i} \rightarrow J_{i}$   is
opposite by the sign to the angular anisotropy in  the
tran\-si\-tions $J_{i} \rightarrow$ \mbox{$J_{i}\!\pm\!1$},
inasmuch as for $J_{i} \geq 1$
\beq{24}
U(J_{i}\,1\,J_{i}\,2,\,J_{i}\,1) < 0,\;\;
U(J_{i}\!\!\pm\!\!1\,1\,J_{i}\,2,\,J_{i}\,1) > 0.
\eeq
This sign variability of Racah function is due to  the
identity
\beq{25}
\sum_{J_{f}}(2J_{f}+1)\,U(J_{f}LJ_{i}Q,J_{i}L)=
\delta_{Q0}(2L+1)(2J_{i}+1)
\eeq
for any values of $L$, $J_{i}$  and $Q$.

   It is interesting that the angular distributions of
$\gamma$  rays  emitted  in  the  series  of   stretched   $e2$
transitions along the yrast line are the same for  all
transitions \cite{16} notwithstanding that the spin-tensors
of nuclear orientation decrease. Indeed, if the $\gamma$  ray
angular distribution in the tran\-si\-tion
$J_{i} \rightarrow$ \mbox{$J_{f}=J_{i}-2$} is
described by the Eq.\ref{17}, thus taking into account Eq.\ref{19}
we obtain for the $\gamma$ ray angular  distribution  in  the
following tran\-si\-tion
$J'_{i}=J_{i}-2 \rightarrow J'_{f}=J_{i}-4$ ($L=2$)
\begin{eqnarray}
&&w(\vec{n}_{\gamma})\!=\!\frac{1}{4\pi}\!\sum_{Q=0,2,4}
(2Q+1)\,C_{L1Q0}^{L1}\,
U(J_{i}\!\!-\!\!4\,L\,J_{i}\!\!-\!\!2\,Q,\,J_{i}\!\!-\!\!2\,L)
\!\cdot \nonumber \\
&& \;\;\;\;\;\;\;\;\;\;\;\;\;\;\;\;\;\;
\cdot U(L\,J_{i}\!\!-\!\!2\,J_{i}\,Q,\,J_{i}\,J_{i}\!\!-\!\!2)
\,\tau_{Q0}(J_{i})\,P_{Q}(\cos \theta_{\gamma}).
\label{26}
\end{eqnarray}
Using the explicit expressions for Racah functions for
$L=2$ and $Q=2,4$ we get
\begin{eqnarray}
&&U(J_{i}\!\!-\!\!4\,L\,J_{i}\!\!-\!\!2\,Q,\,J_{i}\!\!-\!\!2\,L)
\,U(L\,J_{i}\!\!-\!\!2\,J_{i}\,Q,\,J_{i}\,J_{i}\!\!-\!\!2)=
\nonumber \\
&&\;\;\;\;\;\;\;\;\;\;\;\;
=U(J_{i}\!\!-\!\!2\,L\,J_{i}\,Q,\,J_{i}\,L), \label{27}
\end{eqnarray}
therefore the angular distributions  corresponding  to
the se\-quen\-tial transitions
$J_{i} \rightarrow J_{i}-2 \rightarrow J_{i}-4$ coincide.

\section{Estimate of $\gamma$ ray anisotropy}

   A  great  number  of  $e2$  transitions  between  the
yrast-line states of the  fragments  from  spontaneous
fission of\nuc{252}{Cf} was investigated in Ref.\cite{15}.
Among the even-even fragments with high yields
the nuclei \nuc{104}{Mo} and \nuc{144}{Ba}
have  typical  spectra.  The   transitions
$8^{+} \rightarrow 6^{+} \rightarrow 4^{+}
\rightarrow 2^{+} \rightarrow 0^{+}$
correspond to the $\gamma$  ray  energies  606
keV, 520 keV, 369  keV,  and  193  keV  in  the  first
nucleus and 511 keV, 432 keV, 331 keV, and 199 keV  in
the second one. The above discussed "conservation"  of
the  $\gamma$  ray  angular  anisotropy  in   the   stretched
yrast-line transitions gives the  natural  ex\-pla\-na\-tion
of week dependence of angular distributions on  the  $\gamma$
ray ener\-gies found at Ref.\cite{6} (see  also  Ref.\cite{14},
where
the energy dependence of $\gamma$ ray  angular  distributions
was investigated for the binary spontaneous fission of
\nuc{252}{Cf} ).

   Now we  go  to  the  analysis  of  the  differences
between the $\gamma$ ray angular distributions in binary  and
ternary fission. I assume that the shift  of  maximums
from $0^{0}$  and $180^{0}$  to $20-40^{0}$  and $140-160^{0}$
with  respect
to the fission axis is due to the term proportional to
$P_{4}(\cos \theta_{\gamma})$. This requires the high value of
spin-tensor
$\tau_{40}(J_{i})$. To study the sensitivity  of  $\gamma$  ray  angular
distribution to the fragment  alignment  we  take  the
fragment helicity distribution in  the  Gaussian  form
(the spins are aligned perpendicular  to  the  fission
axis)
\beq{28}
\gamma_{K}=\exp(-\frac{K^{2}}{2\sigma^{2}})/
\sum_{K'} \exp(-\frac{K'^{2}}{2\sigma^{2}}),\;\;\;\;
\sum_{K} \gamma_{K} =1.
\eeq

   The spin-tensors $\tau_{20}(J)$ and $\tau_{40}(J)$  of  orientation
of nucleus with spin $J=8$ as functions of  parameter  $\sigma$
are shown in Fig.\ref{Fig.2}. The  populations $\gamma_{K}$
\begin{figure}
\vspace{8cm}
\caption{The spin-tensors $\tau_{20}(J)$ (solid line) and
$\tau_{40}(J)$
(dashed line) of orientation of nucleus with spin  $J\!=\!8$
versus pa\-ra\-me\-ter~$\sigma$ of Gaussian  helicity
distribution.
\label{Fig.2} }
\end{figure}
of  helicity
states are presented in Fig.\ref{Fig.3}
\begin{figure}
\vspace{8cm}
\caption{The populations $\gamma_{K}$  of substates with helicity $K$
for nucleus  with  spin  $J=8$  corresponding  to  $\sigma=0.5$
(circles and solid  line),  $2.5$  (squares  and  dashed
line) and $5.0$ (triangles and dotted line).
\label{Fig.3} }
\end{figure}
for $\sigma=0.5, 2.5 $ and $ 5.0 $.
The  angular  distributions  of  $\gamma$  rays  emitted   in
quadrupole transition $J_{i}=8 \rightarrow J_{f}=6$
from  the  nucleus
whose alignment is determined
by the same values  $0.5$,
$2.5$ and $5.0$ of parameter $\sigma$ are shown
in Fig.\ref{Fig.4} together
\begin{figure}
\vspace{12cm}
\caption{a. The calculated $\gamma$ ray angular distributions in
the transition $J_{i}\!\!=\!\!8\!\rightarrow\!J_{f}\!\!=\!\!6$
corresponding  to  \mbox{$\sigma=0.5$}
(solid line), $2.5$ (dashed line) and $5.0$ (dotted line).
b. The measured in Ref.\protect\cite{6} angular  distributions  of
$\gamma$
rays with energies 540-656~keV from  binary  (crosses)
and ternary (circles) spontaneous fission of \nuc{252}{Cf}.
\label{Fig.4} }
\end{figure}
with the fragment of experimental  data  \cite{6}.  We  see
that the difference between binary and ternary fission
seems to follow from the  difference  between  initial
alignments   of   fission   fragments.   The   angular
distributions in the  subsequent  transitions
$6 \rightarrow 4 \rightarrow 2 \rightarrow 0$
are  the  same  as  presented   in   Fig.\ref{Fig.4}   but   the
distributions  of   populations   differ   from   that
corresponding to $J=8$. Fig.\ref{Fig.5} represents  the  evolution
\begin{figure}
\vspace{13cm}
\caption{The populations $\gamma_{K}$ of substates with helicity $K$
for   nucleus   undergoing    sequential    quadrupole
transitions between the states
$J\!=\!8\;(a) \rightarrow J\!=\!6\;(b) \rightarrow J\!=\!4\;(c)
\rightarrow J\!=\!2\;(d)$  provided  that  the  initial  helicity
distribution   ($J\!=\!8$)   is   of   the   Gaussian   form
corresponding to $\sigma=0.5$.
\label{Fig.5} }
\end{figure}
of  the   initial   Gaussian   helicity   dis\-tri\-bu\-tion
corresponding to $\sigma=0.5$; we  see  that  the  change  of
populations is less than one  might  expect  from  the
decrease of spin-tensors.

\section{Conclusions}

   The data  obtained  in  Ref.\cite{6}  demonstrate
that  $\alpha$
particle emis\-sion in ternary fission does not  destroy
the fragment alignment, but on the  contrary,  ternary
fission strongly correlates  with  high  alignment  of
fission fragments. This result may  be  understood  on
the assumption that the mechanism of  ternary  fission
is closely related with the excitation  of  collective
modes  during  the   saddle-to-scission   stage.   For
example, let us suppose that the $\alpha$  particle  emission
occurs only  if  the  scission  of  bending  fragments
happens just at the moment when  both  fragments  have
the maximal angular velocities; in  this  case  the  $\alpha$
particle emission should correlate with high and  well
aligned fragment spins.

   The proposed mechanism should lead to the  increase
of yield of  fragments  with  high  spins  in  ternary
fission compared with binary one. This assumption  may
be checked by comparison between the  yields  of  high
spin isomers in ternary and binary fission (see, e.g.,
Ref.\cite{17}, where  the  relation  between  the  yields  of
isomers   and   the   initial   fragment   spins   was
established).

   The angular distributions measured in Ref.\cite{6} consist
of the amounts from a great number of  fragments  with
different level schemes. The interpretation of data on
angular distributions of specific $\gamma$ rays emitted  from
even-even fragments (see Ref.\cite{15}) would  be  much  more
reliable. The results  of  calculations  pre\-sen\-ted  in
Fig.\ref{Fig.4}  correspond  really  to  the  transitions  in  a
se\-pa\-ra\-ted even-even fragment.

   Recently  the  enhancement   of   ternary   fission
probability for uranium fissioning isomers  was  found
\cite{18}. In the framework  of  the  suggested  hypothesis
this enhancement may be  caused  by  the  increase  of
probability of excitation of collective modes  at  the
descent stage due  to  some  peculiarities  of  isomer
structure. By the same way the irregularities in total
kinetic  energy  of  fragments,  observed  in  neutron
induced fission near vibrational resonances \cite{19},  may
be explained, but in this case the cor\-re\-la\-tion between
these  irregularities  and  probability   of   ternary
fission should be directly investigated.

   The final  remark  concerns  the  fission  fragment
angular dis\-tri\-bu\-tion. As   it   was   claimed   in
introduction  the  nonaxial  col\-lec\-tive   modes   like
bending or wriggling would distort the $K$  dis\-tri\-bu\-tion
for\-med at the saddle  point.  The  comparison  between
angular dis\-tri\-bu\-tions of  fragments  from  binary  and
ternary  fission  of  aligned  nuclei  may  make  this
distortion evident if really the  role  of  collective
modes is more significant in ternary fission  than  in
binary one.

\end{document}